\documentclass[sigconf]{acmart}
\AtBeginDocument{%
  }

\setcopyright{acmlicensed}
\copyrightyear{2026}
\acmYear{2026}
\acmDOI{XXXXXXX.XXXXXXX}

\acmConference[CHI '26]{CHI Conference on Human Factors in Computing Systems}
  {April 26--30, 2026}{Barcelona, Spain}

\acmISBN{978-1-4503-XXXX-X/26/04}

\usepackage{tabularx}

\begin{document}

\setlength{\abovecaptionskip}{0cm} % 可根据需要调整
\setlength{\belowcaptionskip}{0.1cm} % 可根据需要调整

\title[Co-Designing AI Standardized Patients with Medical Learners]
{``It Talks Like a Patient, But Feels Different'': Co-Designing AI Standardized Patients with Medical Learners}

% -------------------- Authors --------------------
\author{Zhiqi Gao}
\authornote{These authors contributed equally to this work.}
\email{zhiqigao@link.cuhk.edu.cn}
\affiliation{%
  \institution{School of Data Science, The Chinese University of Hong Kong, Shenzhen}
  \city{Shenzhen}
  \state{Guangdong}
  \country{China}
}

\author{Guo Zhu}
\authornotemark[1]
\email{guozhu@link.cuhk.edu.cn}
\affiliation{%
  \institution{School of Artificial Intelligence, The Chinese University of Hong Kong, Shenzhen}
  \city{Shenzhen}
  \state{Guangdong}
  \country{China}
}

\author{Huarui Luo}
\email{2120251953@mail.nankai.edu.cn}
\affiliation{%
  \institution{School of Medicine, Nankai University}
  \city{Tianjin}
  \country{China}
}

\author{Dongyijie Primo Pan}
\email{dpan750@connect.hkust-gz.edu.cn}
\affiliation{%
  \institution{The Hong Kong University of Science and Technology (Guangzhou)}
  \city{Guangzhou}
  \state{Guangdong}
  \country{China}
}

\author{Haoming Tang}
\email{haomingtang@link.cuhk.edu.cn}
\affiliation{%
  \institution{The Chinese University of Hong Kong, Shenzhen}
  \city{Shenzhen}
  \state{Guangdong}
  \country{China}
}

\author{Bingquan Zhang}
\email{zhangkaikai727@gmail.com}
\affiliation{%
  \institution{School of Data Science, The Chinese University of Hong Kong, Shenzhen}
  \city{Shenzhen}
  \state{Guangdong}
  \country{China}
}

\author{Jiahuan Pei}
\email{ppsunrise99@gmail.com}
\affiliation{%
  \institution{Vrije Universiteit Amsterdam}
  \city{Amsterdam}
  \country{Netherlands}
}

\author{Jie Li}
\authornote{Corresponding authors.}
\email{jieli8@mit.edu}
\affiliation{%
  \institution{MIT Media Lab, MIT}
  \city{Cambridge}
  \state{Massachusetts}
  \country{United States}
}

\author{Benyou Wang}
\authornotemark[2]
\email{wangbenyou@cuhk.edu.cn}
\affiliation{%
  \institution{School of Data Science, The Chinese University of Hong Kong, Shenzhen}
  \city{Shenzhen}
  \state{Guangdong}
  \country{China}
}

\renewcommand{\shortauthors}{Gao et al.}

\begin{abstract}
Standardized patients (SPs) play a central role in clinical communication training but are costly, difficult to scale, and inconsistent. Large language model (LLM)–based AI standardized patients (AI-SPs) promise flexible, on-demand practice, yet learners often report that they ``talk like a patient but feel different.'' We interviewed 12 clinical-year medical students and conducted three co-design workshops with them to examine how learners experience SP constraints and what they expect from AI-SPs. We identified six learner-centered needs that are translated into AI-SP design requirements and synthesized a conceptual workflow. Our findings position AI-SPs as tools for deliberate practice and show that instructional usability rather than conversational realism alone drives learner trust, engagement, and educational value.
\end{abstract}




%%%Primo Version Figures
\begin{teaserfigure}
  \includegraphics[width=\textwidth]{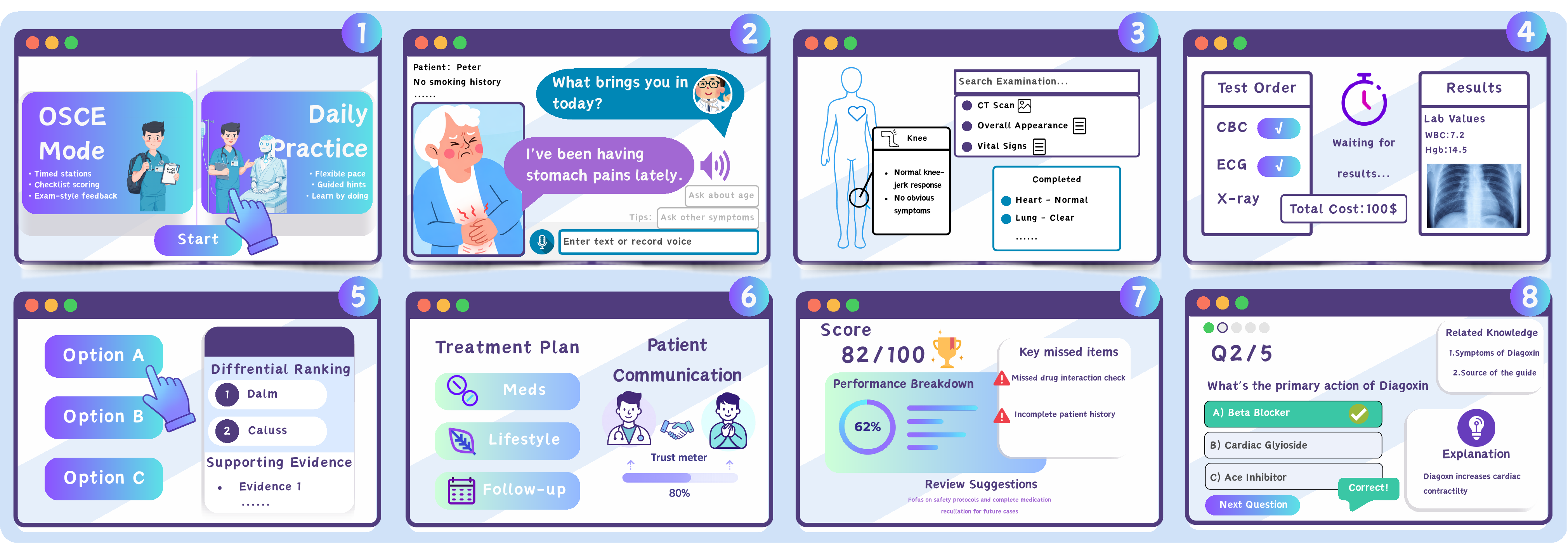}
  \caption{Conceptual workflow of the co-designed AI standardized patient (AI-SP) system. Learners (1) choose OSCE mode or daily practice, then (2) conduct patient history-taking via voice/text with optional prompts, (3) perform guided virtual examinations using a body map and checklist, and (4) order tests and review results with cost cues. Afterward, they (5) rank differential diagnoses with supporting evidence, (6) communicate with the patient and draft a treatment plan, receive (7) in-action performance analytics, and complete (8) a targeted post-session knowledge quiz with explanations and related knowledge expansion.}
  \Description{Eight numbered UI mockups arranged in two rows illustrate a complete training session with an AI standardized patient. Panel 1 is a landing page offering two large options: an OSCE mode and a self-paced daily practice mode. Panel 2 shows a dialogue-based encounter with a patient avatar and an input box for typing or recording voice, plus small hint chips. Panel 3 depicts a virtual physical exam interface with a body silhouette, selectable exam items, and a completion checklist. Panel 4 shows ordering diagnostic tests and viewing results alongside waiting time and total cost. Panel 5 presents differential-diagnosis ranking with a list of candidate diagnoses and supporting evidence. Panel 6 summarizes a treatment plan (meds, lifestyle, follow-up) and a patient-communication area with a trust meter. Panel 7 displays a session score, performance breakdown, and key missed items with review suggestions. Panel 8 shows a short post-session quiz with answer choices, correctness feedback, and related knowledge links.}
  \label{fig:teaser}
\end{teaserfigure}

\maketitle

%% Primo Version
\section{Introduction}

Effective doctor–patient communication is a core clinical competency, shaping information gathering, trust, and patient-centered decision making. Standardized patients (SPs), who are trained actors portraying clinical cases, remain a cornerstone of communication training and OSCE (Objective Structured Clinical
Exam)-style assessment because they provide realistic, repeatable encounters without risking patient safety \cite{barrows1993overview, cleland2009use, nestel2014simulated}. Learners often perceive SPs as offering greater communication realism than other simulation modalities, such as mannequin-based or computer-based virtual patients \cite{meerdink2021comparison}. Despite their educational value, SP programs face persistent practical constraints. Recruiting, training, and scheduling actors is resource-intensive, and performance and feedback quality can vary across sessions \cite{sterz2022manikins}. Such variability challenges standardization and fairness in high-stakes assessments \cite{flanagan2023standardized}, while cost and logistics limit scalability and equitable access \cite{elendu2024impact}. These limitations motivate growing interest in AI-supported alternatives that can provide on-demand, repeatable, and configurable communication practice.

Earlier digital simulated patients (e.g., virtual patients) improved accessibility but relied on scripted dialogue trees or rule-based interactions, limiting conversational flexibility and clinical flow \cite{cook2010computerized, cook2009virtual}. Large language models (LLMs) now support more open-ended, context-aware patient dialogue \cite{brown2020language}, motivating LLM-driven AI standardized patients (AI-SPs) for scalable simulation, automated feedback, and immersive interaction \cite{holderried2024generative, hicke2025medsimai, zhu2025vaps}. However, educational adoption depends not only on linguistic fluency but also on learner experience: prior evaluations report an ``empathy gap'' and the absence of non-verbal cues critical for rapport \cite{borg2024creating, tu2024towards}. Persona drift and hallucinated clinical details further raise safety and validity concerns \cite{ferrario2024role}, and even empathy-oriented systems struggle to reproduce the relational nuance learners expect \cite{steenstra2025simpatient}. 

Given these gaps, our \textbf{Research Question (RQ)} asks: \textit{How do clinical-year medical learners experience SP-based communication training and co-envision AI-SP designs that balance standardization, clinical realism, and relational authenticity?} We address this through interviews and co-design workshops with clinical-year medical students, synthesizing learner-identified needs and design proposals into a set of AI-SP requirements.

%%Primo Version RW:
\section{Related Work}

\subsection{Learners' Experience in Clinical Communication Training}

Clinical communication training goes beyond diagnostic reasoning or linguistic fluency to include relational and emotional labor, such as building rapport, conveying empathy, and regulating affect in high-stakes interactions. Medical education research frames clinical empathy as an active process of managing one’s own emotions while responding to patients’ verbal and non-verbal cues \cite{larson2005clinical, shapiro2011empathy}. Although central to patient trust and care quality, these competencies remain difficult to teach or assess through procedural checklists alone.

Standardized patients (SPs) are widely studied for supporting learners' relational and emotional competencies. By enabling embodied, socially responsive interaction, SPs allow learners to practice rapport-building, empathy, and affect regulation beyond scripted or procedural tasks \cite{barrows1993overview, cleland2009use, nestel2014simulated, rutherford2024use}. From learners' perspectives, SP quality is tied more to interactional flow, emotional attunement, and feeling understood than to informational accuracy alone \cite{shapiro2011empathy}. However, SP-based training faces scalability constraints, including cost, scheduling, and variability in performance and feedback \cite{sterz2022manikins, flanagan2023standardized, elendu2024impact}. Earlier virtual patients addressed access but relied on scripted or rule-based dialogue, limiting conversational flexibility and encounter dynamics \cite{cook2009virtual, cook2010computerized}.

\subsection{LLM-Driven AI Standardized Patients }
More recently, LLMs have enabled more open-ended and context-aware simulated dialogue \cite{brown2020language}, motivating the development of AI-driven standardized patients (AI-SPs) that support diverse personas, automated feedback, and immersive interaction, including VR-based embodiment \cite{holderried2024generative, hicke2025medsimai, zhu2025vaps}. Despite these advances, evaluations consistently report an ``empathy gap'' that can generate fluent patient dialogue but feel relationally different to learners, often attributed to limited affective responsiveness \cite{borg2024creating, tu2024towards}. Even systems designed to scaffold empathy through feedback continue to struggle to reproduce the relational nuance learners expect from clinical encounters \cite{steenstra2025simpatient}. Persona inconsistency and hallucinated clinical details further raise pedagogical and safety concerns for training validity \cite{ferrario2024role}.

Previous work suggests that challenges in AI-SP adoption are not solely technical but experiential, motivating learner-centered inquiry into how AI-SPs should balance standardization, clinical realism, and relational authenticity.

\section{Method}

\subsection{Participants and Study Design}
\textbf{Participants.} We recruited 12 clinical-year medical students (Years 4–6) from two medical schools, all with prior experience using human SPs (e.g., communication training and/or OSCE preparation). Participants had diverse clinical rotation backgrounds. Recruitment used institutional mailing lists and snowball sampling \cite{parker2019snowball}. Participation was voluntary, unrelated to academic evaluation, and compensated with 20 USD. Participant demographics are provided in Appendix~A.

\noindent\textbf{Study Design \& Procedure.} We conducted a two-phase qualitative study.
\textbf{Phase 1: Semi-structured interviews.} We conducted remote interviews with 12 participants via video conferencing ($\approx$45 minutes), focusing on experiences with SPs, including perceived realism, emotional and cognitive load, breakdown moments, and feedback practices. Interviews were audio-recorded. \textbf{Phase 2: Co-design workshops.} We ran three in-person co-design workshops with the same participants (4 per workshop; $\approx$75 minutes) following established co-design practices \cite{sanders2008cocreation}. Participants individually sketched an ``ideal AI-SP'' reflecting usage context, interaction modes, realism/affect, and feedback, then rotated and annotated sketches across two rounds, followed by group discussion. All artifacts were photographed and discussions audio-recorded.

\subsection{Ethics and Data Analysis}
The study was approved by the institutional ethics review board (IRB). All participants provided written informed consent. Transcripts and artifacts were de-identified, and only anonymized quotations (P1–P12) are reported. Audio recordings, transcripts, and photos were stored on access-controlled drives and used solely for research purposes.
 % TODO: add IRB/ethics approval/exemption statement per your institution.

Our analysis followed a two-stage logic. Phase 1 interview data were analyzed using reflexive thematic analysis \cite{braun2006thematic, braun2019reflexive}. We conducted line-by-line initial coding to capture incidents and learner concerns \cite{charmaz2006gt}. We then iteratively consolidated codes into candidate themes through team discussion. Phase 2 workshop data were analyzed deductively using the Phase 1 needs as an analytic framework. Design solutions proposed by participants were first extracted as candidate features and then systematically mapped onto the interview-derived needs. Features addressing the same underlying need were consolidated into AI-SP design requirements. Throughout the analysis, the team met regularly to compare interpretations and resolve disagreements through discussion~\cite{braun2019reflexive}.

\section{Results}
Phase 1 interviews identified six themes reflecting needs in SP-based training, while Phase 2 co-design workshops translated these needs into AI-SP design requirements. Synthesizing both phases, we articulate a conceptual workflow that operationalizes learner-identified needs into usable AI-SP system requirements. Table~\ref{tab:themes} summarizes each theme as a needs–solutions–requirements chain with representative quotes.

\subsection{Six themes : Learner-Identified Needs in SP-Based Training.} 

Across interviews, six themes (T1–T6) captured recurring needs and tensions in SP-based training (Table~\ref{tab:themes}). Learners first described a mismatch between training intent and simulation fidelity, noting that SPs serve different purposes at different stages (e.g., exam preparation vs.\ exploratory learning) and should therefore be structured and evaluated differently (\textbf{T1}). They also questioned clinical realism in human SPs: despite authentic performances, biased or predictable information release can turn encounters into ``games'' rather than investigations (\textbf{T2}). Learners further highlighted limits on what can be safely practiced with human actors, especially physical examination, which restricted evidence gathering and reinforced purely conversational interaction (\textbf{T3}). They described affective barriers in SP sessions, where limited in-action support and delayed feedback increased anxiety and obscured communication breakdowns (\textbf{T4}), and expressed frustration with feedback that arrived too late or lacked situational grounding (\textbf{T5}). Finally, they emphasized that communication training must address emotional labor and patient education, not only diagnostic correctness, requiring practice with varied emotional reactions during explanation, negotiation, and shared decision making (\textbf{T6}).

Together, these themes frame SP training challenges as experiential tensions involving realism, support, and relational work, rather than purely logistical or technical issues.

\subsection{Co-design Outcomes: Mapping Learner Needs to AI-SP Design Requirements}

In co-design workshops, participants translated the interview-identified needs into six AI-SP design requirements (D1–D6). \textbf{D1: Goal-aligned fidelity modes.} Participants proposed explicit mode selection (e.g., OSCE vs.\ daily practice), with each mode defining distinct goals, realism targets, and evaluation criteria, supporting exam standardization in some contexts and exploratory reasoning in others. \textbf{D2: Policy-driven information release.} To prevent encounters from becoming predictable ``games,'' learners suggested defining clear rules for when clinical information is revealed: salient perceptual cues (e.g., visible jaundice) are always present, while deeper information remains inquiry-dependent, preserving both fairness and realism. \textbf{D3: Multimodal evidence interaction.} To address limits of human SPs, participants proposed extending interaction beyond dialogue through examination interfaces, sensory cues, and test-ordering tools, enabling practice of evidence gathering alongside communication. \textbf{D4: Hybrid input as a reliability layer.} Learners anticipated breakdowns and anxiety in purely voice-based interactions. They emphasized hybrid input strategies, in which voice remains the primary interaction mode but text or keyword input and responsive hints are available as backups when breakdowns occur, helping learners continue the interaction and reducing stress. \textbf{D5: Learner-controlled scaffolding and dual-loop feedback.} Participants reframed AI-SPs as tools for deliberate practice rather than summative assessment. They wanted learner-controlled scaffolding, such as optional prompts and adjustable difficulty, so they could move from coached practice to exam-like practice. They also emphasized dual-loop feedback: lightweight cues during the encounter (e.g., changes in perceived trust or affect), followed by structured post-session debriefs that reconstruct missed questions and connect errors to relevant knowledge. \textbf{D6: Controllable affect and relational variability.} \textbf{R6: Controllable affective variability.} Finally, learners proposed selectable patient personalities and emotional responses, with replayable scenarios, to deliberately practice explanation, negotiation, and shared decision making while supporting reflection.

\subsection{Synthesis: From Learner Needs to Conceptual AI-SP Workflows}

Synthesizing Phases 1 and 2, we derived a conceptual AI-SP workflow that integrates learner needs with co-designed solutions. As illustrated in Fig.~\ref{fig:teaser}, the workflow emphasizes (1) goal-aligned fidelity selection through explicit mode choice (Panel~1), (2) multimodal evidence interaction layered onto conversation via voice/text history-taking, guided examinations, and test ordering (Panels~2–4), and (3) learner-controlled scaffolding that supports progression from coached practice to stress-testing across key interaction stages (Panels~2–6). Learners further emphasized (4) dual-loop feedback, combining in-action relational and performance cues with structured post-session analytics and knowledge consolidation (Panels~6–8), and (5) affective variability treated as a configurable training dimension, enabling practice of emotional labor without undermining reflective control (Panels~2,~6). This workflow demonstrates how AI-SP systems can balance standardization, clinical realism, and relational authenticity, directly addressing our research question by grounding design requirements in learners' lived experiences and co-envisioned practices.

\begin{table*}[t]
\centering
\scriptsize
\setlength{\tabcolsep}{3pt}
\renewcommand{\arraystretch}{1.12}
\begin{tabularx}{\textwidth}{
  >{\raggedright\arraybackslash}p{0.16\textwidth}
  >{\raggedright\arraybackslash}X
  >{\raggedright\arraybackslash}X
  >{\raggedright\arraybackslash}p{0.30\textwidth}}
\toprule
\textbf{Themes} & \textbf{Interview-identified needs (quotes)} & \textbf{Co-design proposed solutions (quotes)} & \textbf{AI-SP requirements} \\
\midrule

\textbf{T1: Context-Dependent Fidelity \& Training Goals}
&
\textit{``SP usage varies by context… one path for exam standardization, another for comprehensive clinical thinking.''} (P1)
&
\textit{``Undergraduates need more knowledge reinforcement… [postgraduate medical] residents need scenarios that continue until the `patient' is satisfied.''} (WS-A)
&
Support goal- and stage-specific modes (e.g., OSCE vs.\ practice), with adjustable fidelity and evaluation criteria.
\\
\midrule

\textbf{T2: Standardization vs.\ Clinical Realism}
&
\textit{``[SP actors] answer too easily... or become deliberately vague.''} (P12)
&
\textit{``Jaundice should be immediately visible... rather than needing to ask.''} (WS-B)
&
Implement policy-driven information release (case-consistent and inquiry-dependent), while providing perceptual cues that preserve clinical realism.
\\
\midrule

\textbf{T3: Multimodal Interaction \& Evidence Gathering}
&
\textit{``You can't risk puncturing an artery instead of a vein on a real person.''} (P1)
&
\textit{``3D digital human… clicking body parts triggers pain or tenderness.''} (WS-C) \newline
\textit{``Allow text input… a keyword like `cough' triggers hints or autocomplete.''} (WS-A)
&
Go beyond text chat by adding multimodal ``sensory proxies'' (visual signs, exam interactions, sounds) and hybrid input (voice-first with text fallback) to reduce interaction breakdowns and learner anxiety.
\\
\midrule

\textbf{T4: Scaffolding \& Adaptive Difficulty}
&
\textit{``Different levels of difficulty… different modes with different difficulties.''} (P4)
&
\textit{``Friendly Mode... offering prompts if I struggle.''} (WS-B)
&
Provide learner-controlled scaffolding and adjustable difficulty (e.g., coaching mode vs.\ stress-test mode) to support deliberate practice.
\\
\midrule

\textbf{T5: Immediate vs.\ Summative Feedback Loops}
&
\textit{``I prefer knowing… the moment it happened.''} (P5)
&
\textit{``After the session… a `cheating paper' that links mistakes to textbook references.''} (WS-C)
&
Offer dual-loop feedback: in-action cues (e.g., trust meter / facial-expression shifts) plus a structured post-session debrief that reconstructs missed items and reasoning.
\\
\midrule

\textbf{T6: Affective Computing \& the ``Human'' Variable}
&
\textit{``[There should be] patients with different personalities… and different backgrounds… more in line with real-world scenarios.''} (P8)
&
\textit{``Randomized personalities… elderly grandma… a child… crying or throwing a tantrum.''} (WS-A)
&
Model affect as a controllable training variable (persona and emotion profiles) to practice explanation, negotiation, and shared decision making, with replayable debrief.
\\
\bottomrule
\end{tabularx}
\caption{Six themes linking interview-identified needs to co-designed AI-SP requirements (WS-A/B/C denote workshop groups).}
\label{tab:themes}
\end{table*}

\section{Discussion}

% Our research examined how clinical-year learners experience limitations in SP-based communication training and how they envision AI-SP systems that better balance standardization, clinical realism, and relational authenticity. Rather than viewing the commonly reported “empathy gap” in AI-SPs as a problem of emotional expression or language quality alone \cite{borg2024creating, tu2024towards, steenstra2025simpatient}, learners reframed it as a problem of instructional clarity. 

Our research highlighted three qualities that make simulated encounters useful for learning: \textbf{controllability} (what can vary and for what purpose), \textbf{observability} (what makes learning progress visible), and \textbf{learnability} (how feedback helps improvement during and after practice). This perspective shifts attention away from whether AI can simply ``act like'' a patient, toward how AI-based simulations can be designed as transparent, reliable tools for clinical training.

\subsection{From Conversational Realism to Instructional Usability}

Prior work shows that early virtual patients scale but feel scripted, recent LLM-based SPs remain conversationally fluent yet still feel relationally inauthentic to learners \cite{nestel2014simulated, holderried2024generative, hicke2025medsimai, zhu2025vaps}. Our findings indicate that learners did not judge realism by how natural the dialogue sounded, but by whether the simulation matched what they were practicing. An AI-SP meant for OSCE exam rehearsal, for example, should behave differently from one intended for open-ended clinical exploration or skills practice. Conversational realism alone was therefore not enough. What mattered was whether the system made its purpose, limits, and evaluation rules clear. Explicit mode selection helps learners know what kind of performance was expected and reducing uncertainty and mistrust, aligning with prior HCI guidance on transparency and user control in high-stakes AI systems \cite{amershi2019guidelines}.

\subsection{Evidence-Gathering Gap in Communication Training}

Our work also revealed the challenges in the evidence-gathering gap. Prior work on SPs tends to treat communication as primarily conversational, implicitly separating ``talking'' from ``doing.''~\cite{barrows1993overview, cleland2009use, nestel2014simulated, rutherford2024use} Learners in our study emphasized that effective communication is inseparable from evidence work: asking the right questions, performing examinations, and interpreting cues. Human SPs limit this integration because physical examination are constrained by safety and practicality. Learners argued that AI-SPs can intervene not by perfectly simulating bodies, but by providing instrumented surrogates for evidence gathering. Multimodal interaction (e.g., guided body maps, sensory cues, or interfaces for ordering and reviewing clinical tests) was framed as a way to practice how communication and evidence co-evolve, rather than as a pursuit of sensory realism.

\subsection{Positioning AI-SPs Within a Training Ecology}

Our findings position AI-SPs not as replacements for human SPs, but as complementary infrastructure within a broader training ecology. Human SPs remain essential for high-stakes assessment, embodied presence, and nuanced social interaction. AI-SPs, by contrast, can offer repeatable practice, configurable difficulty, structured evidence work, and interpretable feedback between limited actor-led sessions. This layered view aligns with recent evaluations of LLM-based SPs that note feasibility alongside persistent concerns about realism and validity \cite{cross2025chatgptsp, li2026llmvp}. This also highlights a core challenge in healthcare AI: strong models alone are not enough. AI systems must be translated into training tools that fit clinical practice, support learning goals, and make sense to learners' everyday experiences \cite{andersen2023hcaihealthcare}.

\subsection{Limitations and Future Work}
This study is a learner-centered qualitative and co-design investigation and does not evaluate learning outcomes or educator-facing constraints such as grading validity or curriculum integration. Future work should implement AI-SP prototypes aligned with these design specifications and evaluate them in authentic settings (e.g., OSCE preparation). In addition, communication norms and educational expectations vary across cultures, replication across institutions and multi-stakeholder co-design with educators and SP trainers will be essential before generalizing these claims.

\section{Conclusion}
This study shows that learners experience limitations in current SP-based training not only as technical issues, but as gaps in instructional clarity and support. By grounding AI-SP design in learners' experiences, we outline how AI-SPs can complement human SPs through deliberate practice, structured evidence work, and interpretable feedback. Our findings position AI-SPs as instructional infrastructure rather than conversational substitutes.

\section*{Generative AI Use Disclosure}
Generative AI was used only for language editing during writing. All analysis, interpretation, and claims were conducted and verified by the authors.

\bibliographystyle{plain}
\newpage
\bibliography{reference}

\begin{thebibliography}{10}

\bibitem{amershi2019guidelines}
Saleema Amershi, Daniel Weld, Mihaela Vorvoreanu, Adam Fourney, Besmira Nushi, Penny Collisson, Jina Suh, Shamsi Iqbal, Paul~N. Bennett, Kori Inkpen, Jaime Teevan, Ruth Kikin-Gil, and Eric Horvitz.
\newblock Guidelines for human-ai interaction.
\newblock In {\em Proceedings of the 2019 CHI Conference on Human Factors in Computing Systems}. ACM, 2019.

\bibitem{andersen2023hcaihealthcare}
Tariq~Osman Andersen, Francisco Nunes, Lauren Wilcox, Enrico Coiera, and Yvonne Rogers.
\newblock Introduction to the special issue on human-centred ai in healthcare: Challenges appearing in the wild.
\newblock {\em ACM Transactions on Computer-Human Interaction}, 30(2), 2023.

\bibitem{barrows1993overview}
Howard~S Barrows.
\newblock An overview of the uses of standardized patients for teaching and evaluating clinical skills.
\newblock {\em Academic Medicine}, 68(6):443--451, 1993.

\bibitem{borg2024creating}
Anton Borg, Ioannis Parodis, and Gabriel Skantze.
\newblock Creating virtual patients using robots and large language models: A preliminary study with medical students.
\newblock In {\em Companion of the 2024 ACM/IEEE International Conference on Human-Robot Interaction (HRI '24)}, pages 273--277, 2024.

\bibitem{braun2006thematic}
Virginia Braun and Victoria Clarke.
\newblock Using thematic analysis in psychology.
\newblock {\em Qualitative Research in Psychology}, 3(2):77--101, 2006.

\bibitem{braun2019reflexive}
Virginia Braun and Victoria Clarke.
\newblock Reflecting on reflexive thematic analysis.
\newblock {\em Qualitative Research in Sport, Exercise and Health}, 11(4):589--597, 2019.

\bibitem{brown2020language}
Tom~B Brown, Benjamin Mann, Nick Ryder, et~al.
\newblock Language models are few-shot learners.
\newblock {\em arXiv preprint arXiv:2005.14165}, 2020.

\bibitem{charmaz2006gt}
Kathy Charmaz.
\newblock {\em Constructing Grounded Theory: A Practical Guide Through Qualitative Analysis}.
\newblock SAGE, 2006.

\bibitem{cleland2009use}
Jennifer~A Cleland, Kazuya Abe, and Jan-Joost Rethans.
\newblock The use of simulated patients in medical education: Amee guide no 42.
\newblock {\em Medical Teacher}, 31(6):477--486, 2009.

\bibitem{cook2010computerized}
David~A Cook, Patricia~J Erwin, and Marc~M Triola.
\newblock Computerized virtual patients in health professions education: a systematic review and meta-analysis.
\newblock {\em Academic Medicine}, 85(10):1589--1602, 2010.

\bibitem{cook2009virtual}
David~A Cook and Marc~M Triola.
\newblock Virtual patients: a critical literature review and proposed next steps.
\newblock {\em Medical Education}, 43(4):303--311, 2009.

\bibitem{cross2025chatgptsp}
Joseph Cross, Tarron Kayalackakom, Raymond~E. Robinson, Andrea Vaughans, Roopa Sebastian, Ricardo Hood, Courtney Lewis, Sumanth Devaraju, Prasanna Honnavar, Sheetal Naik, Jillwin Joseph, Nikhilesh Anand, Abdalla Mohammed, Asjah Johnson, Eliran Cohen, Teniola Adeniji, Aisling~Nnenna Nnaji, and Julia~Elizabeth George.
\newblock Assessing chatgpt's capability as a new age standardized patient: Qualitative study.
\newblock {\em JMIR Medical Education}, 11:e63353, 2025.

\bibitem{elendu2024impact}
Chukwuebuka Elendu, Daniella~C Amaechi, et~al.
\newblock The impact of simulation-based training in medical education: A review.
\newblock {\em Medicine}, 103(27):e38813, 2024.

\bibitem{ferrario2024role}
Andrea Ferrario, Jana Sedlakova, and Manuel Trachsel.
\newblock The role of humanization and robustness of large language models in conversational artificial intelligence for individuals with depression: A critical analysis.
\newblock {\em JMIR Mental Health}, 11:e56569, 2024.

\bibitem{flanagan2023standardized}
Olivia~L Flanagan and Kathleen~M Cummings.
\newblock Standardized patients in medical education: A review of the literature.
\newblock {\em Cureus}, 15(7):e42027, 2023.

\bibitem{hicke2025medsimai}
Yann Hicke, Jadon Geathers, Niroop Rajashekar, et~al.
\newblock Medsimai: Simulation and formative feedback generation to enhance deliberate practice in medical education.
\newblock {\em arXiv preprint arXiv:2503.05793}, 2025.

\bibitem{holderried2024generative}
Friederike Holderried, Christian Stegemann-Philipps, Lea Herschbach, et~al.
\newblock A generative pretrained transformer (gpt)-powered chatbot as a simulated patient to practice history taking: prospective, mixed methods study.
\newblock {\em JMIR Medical Education}, 10(1):e53961, 2024.

\bibitem{larson2005clinical}
Eric~B Larson and Xin Yao.
\newblock Clinical empathy as emotional labor in the patient-physician relationship.
\newblock {\em Jama}, 293(9):1100--1106, 2005.

\bibitem{li2026llmvp}
Dongliang Li and Syaheerah~Lebai Lutfi.
\newblock Large language model-based virtual patient systems for history-taking in medical education: Comprehensive systematic review.
\newblock {\em JMIR Medical Informatics}, 14:e79039, 2026.

\bibitem{meerdink2021comparison}
Margaretha Meerdink and Javaid Khan.
\newblock Comparison of the use of manikins and simulated patients in a multidisciplinary in situ medical simulation program for healthcare professionals in the united kingdom.
\newblock {\em Journal of Educational Evaluation for Health Professions}, 18:8, 2021.

\bibitem{nestel2014simulated}
Debra Nestel and Margaret Bearman.
\newblock Simulated patient methodology in health professional education.
\newblock pages 1--10, 2014.

\bibitem{parker2019snowball}
Charlie Parker, Sam Scott, and Alistair Geddes.
\newblock Snowball sampling.
\newblock {\em SAGE research methods foundations}, 2019.

\bibitem{rutherford2024use}
Tonya Rutherford-Hemming, Alaina Herrington, and Thye~Peng Ngo.
\newblock The use of standardized patients to teach communication skills—a systematic review.
\newblock {\em Simulation in Healthcare}, 19(1S):S122--S128, 2024.

\bibitem{sanders2008cocreation}
Elizabeth B.-N. Sanders and Pieter~Jan Stappers.
\newblock Co-creation and the new landscapes of design.
\newblock {\em CoDesign}, 4(1):5--18, 2008.

\bibitem{shapiro2011empathy}
Johanna Shapiro.
\newblock 16 the paradox of teaching empathy in medical education.
\newblock {\em Empathy: From bench to bedside}, page 275, 2011.

\bibitem{steenstra2025simpatient}
Ian Steenstra, Farnaz Nouraei, and Timothy Bickmore.
\newblock Scaffolding empathy: Training counselors with simulated patients and utterance-level performance visualizations.
\newblock In {\em Proceedings of the {CHI} Conference on Human Factors in Computing Systems}, CHI '25, Yokohama, Japan, 2025. Association for Computing Machinery.

\bibitem{sterz2022manikins}
Janina Sterz, Nils Gutenberger, Maria~Cristina Stefanescu, et~al.
\newblock Manikins versus simulated patients in emergency medicine training: a comparative analysis.
\newblock {\em European Journal of Trauma and Emergency Surgery}, 48(5):3793--3801, 2022.

\bibitem{tu2024towards}
Tao Tu, Anil Palepu, Mike Schaekermann, et~al.
\newblock Towards conversational diagnostic ai.
\newblock {\em arXiv preprint arXiv:2401.05654}, 2024.

\bibitem{zhu2025vaps}
Xiuqi~Tommy Zhu, Heidi Cheerman, Minxin Cheng, Sheri~R. Kiami, Leanne Chukoskie, and Eileen McGivney.
\newblock Designing {VR} simulation system for clinical communication training with {LLMs}-based embodied conversational agents.
\newblock In {\em Extended Abstracts of the {CHI} Conference on Human Factors in Computing Systems}, CHI EA '25, Yokohama, Japan, 2025. Association for Computing Machinery.

\end{thebibliography}

\clearpage
\onecolumn
\appendix
\section{Appendix A: Participant Overview}
\begin{table}[H]
\centering
\small
\setlength{\tabcolsep}{3pt} % 列间距更紧（你现在看着空就是这儿）
\renewcommand{\arraystretch}{1.10}

\begin{tabular}{@{}l l l l l@{}}
\hline
\textbf{ID} & \textbf{Recruitment site} & \textbf{Education level} & \textbf{Training track} & \textbf{Gender} \\
\hline
P01 & MS-A & Graduate & Anesthesiology & Female \\
P02 & MS-B & Undergraduate & Clinical Medicine & Male \\
P03 & MS-A & Graduate & Surgery & Male \\
P04 & MS-B & Undergraduate & Clinical Medicine & Female \\
P05 & MS-B & Undergraduate & Clinical Medicine & Male \\
P06 & MS-B & Undergraduate & Clinical Medicine & Female \\
P07 & MS-B & Undergraduate & Clinical Medicine & Male \\
P08 & MS-B & Undergraduate & Clinical Medicine & Female \\
P09 & MS-B & Undergraduate & Clinical Medicine & Female \\
P10 & MS-B & Undergraduate & Clinical Medicine & Male \\
P11 & MS-A & Graduate & Surgery & Female \\
P12 & MS-A & Graduate & Surgery & Male \\
\hline
\end{tabular}

\caption{Participant overview. MS-A and MS-B denote two anonymized medical schools (institutions not reported for privacy). All participants were clinical-year medical students with prior standardized patient (SP) experience.}
\label{tab:participants}
\end{table}

\end{document}